\documentclass[fleqn,12pt,twoside]{article}
\usepackage{espcrc1}

\usepackage{graphicx}

\newcommand{ \be }{\begin{equation}}
\newcommand{ \ee }{\end{equation}}
\newcommand{ \bea }{\begin{eqnarray}}
\newcommand{ \eea }{\end{eqnarray}}
\newcommand{ \la }{\langle}
\newcommand{ \ra }{\rangle}
\newcommand{ \eps }{{\varepsilon}}
\newcommand{ \pt }{{$p_t$ }}
\newcommand{ \vtpt }{{v_2(p_t) }}
\newcommand{ \snn }{{$\sqrt{s_{NN}} $ }}

\title{Anisotropic flow}

\author{S.A. Voloshin\address{Department of Physics and Astronomy\\ 
        Wayne State University, Detroit, U.S.A.}}
\begin{document}

\maketitle

\begin{abstract}
Recent experimental results on directed and elliptic flow, theoretical
developments, and new techniques for anisotropic flow
analysis are reviewed.  
\end{abstract}

\section{Introduction. Flow adventures from Monterey to Nantes.}

From the point of view of anisotropic flow
this conference 
is quite remarkable. 
Elliptic flow has been fully recognized as an important
observable providing information on the early stages of
heavy ion collisions. 
An increased attention to anisotropic flow in recent years 
has resulted in a significant improvement in all three components of the field:
techniques and methods, analysis and presentation of the data, and
theoretical understanding of the underlying physics.  
For a long time the anisotropic flow measurements suffered large
systematic errors related to the so called ``non-flow'' contribution,
the contribution due to 
the azimuthal correlations not related 
to the orientation of the reaction plane.
At this Quark Matter conference, for the first time, 
we discuss the results mostly free of this uncertainty, 
using results obtained 
with 3-, 4-, and even more particle correlations.
Elliptic flow up to $p_t \sim 10\;-\;12$~GeV/c, 
the ``wiggle'' in directed flow, $v_2(p_t)$ for a number of identified
particles up to \pt$\sim$3-4~GeV/c, are only a few other results 
presented at this conference.

The starting point for many recent  developments 
was the Quark Matter '95 conference at Monterey.
Then, the E877 Collaboration reported the first observation of
anisotropic flow in ultra-relativistic nuclear collisions
at the BNL  AGS~\cite{zhang}, the discovery, which 
M. Gyulassy in his concluding remarks
called the ``major highlight of the year''.
That analyses was done  with the use of a new
approach~\cite{zhang-voloshin}, the very notations of which, like
$v_1$ and  $v_2$, are a common language now. 
The E877 Collaboration also presented the results from the first attempts
to perform HBT interferometry of the anisotropic source, one of the
important directions in today's HBT analysis. 
At the same conference,
J.-Y. Ollitrault, foreseeing future precise measurements,
emphasized the importance of the understanding and
measuring of the non-flow correlations. 
It was pointed out that flow is a collective phenomena and
in order to disentangle it from other contributions to particle
azimuthal correlations one has to exploit multi-particle correlations
(cumulants).

Anisotropic flow is defined as azimuthal asymmetry in
particle distribution with respect to the reaction plane (the plane
spanned by the beam direction and the impact parameter).
It is called {\em flow} for it is a collective phenomena, 
but it does not necessarily imply {\em hydrodynamic} flow.
It is commonly characterized using
Fourier decomposition of the azimuthal distributions~\cite{method}. 
Then the first
harmonic Fourier coefficient, $v_1$, describes {\em directed} flow, 
and the second harmonic coefficient, $v_2$, 
corresponds to {\em elliptic} flow; 
non-zero higher harmonics can be also present in the distribution. 

The two reasons for anisotropic flow are the original asymmetry in the
configuration space (non-central collisions !) and rescatterings.  
In a case of elliptic
flow the initial ``ellipticity'' of the overlap zone is usually
characterized  by the quantity 
$\eps=(\la y^2 - x^2 \ra) /(\la y^2 + x^2 \ra$, assuming the
reaction plane being the $xz$-plane. 
With the system expansion the spatial anisotropy decreases. 
This is the reason for the high sensitivity of elliptic flow to the
evolution of the system at very early times~\cite{Sorge97,mpc},
2--5~fm/c, of the order of the size of the system.

\section{Directed flow}

Three major results reported at this conference regarding
directed flow were: the understanding of the role of momentum
conservation~\cite{wetzler,borghini} 
(due to lack of space I do not discuss it here, for details,
see~\cite{MomConservation}), 
the use of  3-particle correlations in order to suppress
the non-flow contributions~\cite{borghini}, 
and the first results on directed
flow of protons close to mid-rapidity~\cite{wetzler}. 
The latter is related
to earlier predictions of the so called ``wiggle'' in the dependence of
$v_1(y)$~\cite{thirdcomponent,wiggle}. 

{\em Three-particle correlations}, discussed in the talk of Borghini et
al.~\cite{borghini}, relies on the observed strong elliptic flow
at mid-rapidity and permits directed flow measurements, which are
 much less affected by non-flow correlations. If in the usual two-particle
correlator 
\be
\la e^{i(\phi_1 -\phi_2)} \ra \approx v_1^2 +\delta_1,
\ee
the non-flow contribution, $\delta_1 \sim 1/N_{tot}$, 
where $N_{tot}$  is the total number of particles,
then the 3-particle correlation approach gives
\be
\la e^{i(\phi_1 +\phi_2-2\phi_3)} \approx v_1^2 v_2 +\delta_1^2.
\label{three}
\ee
The relative suppression of the non-flow contribution in Eq.~2
compared to Eq.~1 is of the order
of $1/(v_2 N_{tot})$, which for large elliptic flow could be 
significant.  
Mathematically, one can trace that in  
Eq.~\ref{three}, the contribution of any
correlations projected onto the reaction plane and the plane
perpendicular to the reaction plane enter with opposite signs. 
The non-flow contributions, by definition, are the correlations not
dependent on the relative orientation with respect to the reaction
plane, and as such just cancel out from the final result.

The so called {\em ``wiggle'' in directed flow}, the three times change in the
sign of directed flow in the midrapidity region, had been predicted a few
years ago. The origin of this wiggle is somewhat different in
different models, but any of the explanations would be very intriguing and
interesting. 
In the ``third flow component'' picture~\cite{thirdcomponent}, 
it results from the expansion of the initially tilted disk of the compressed
matter. 
It appeared that the effect is present only if QGP is produced. 
Thus the wiggle might be a signature of the QGP production.
The second explanation of the wiggle relies on the non-complete stopping
of nucleons during the collisions, which results in the correlation of
the position of the nucleon in the transverse plane and their
rapidity. 
Combining this with a subsequent radial expansion one gets the
wiggle~\cite{wiggle}.
The first results have been presented by the NA49
Collaboration~\cite{wetzler} (see Fig. 1):
\begin{figure}
\centerline{
\includegraphics[height=8cm]{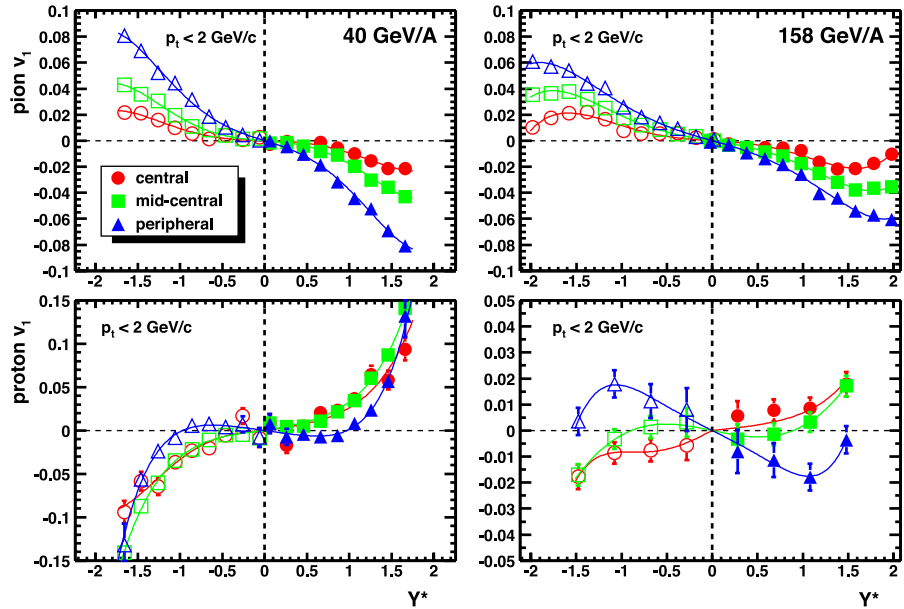}
}
\vspace*{-8mm}
\caption{Directed flow of pions and protons at beam energies of 40$A$ and
160$A$ GeV. Note the different scales in the two proton flow graphs.}
\label{fv1}
\end{figure}
One can see a clear indication of the change of sign of directed flow
of protons as a function of rapidity for the most peripheral events. 
It would be extremely interesting
to see the analogous results from experiments at RHIC.
Detailed model predictions for the energy dependence
of directed flow of both pions and baryons would also be very interesting.

\section{Elliptic flow}

There have been many new results and ideas presented at this
conference. I will only briefly mention the most important (from my
point of view).

\subsection{Comparing the results from different experiments}

Anisotropic flow results at this conference have been presented by
many collaborations~\cite{wetzler,borghini,slivova,esumi,manly}. 
As always in such a case the question appears if these results are
consistent. 
There are obvious difficulties to perform a fair comparison:
different collaborations use different centralities.
Nevertheless, I would conclude that  the data where
comparable are consistent (for the figures see the slides of my
talk).

The second question often asked during the conference 
concerns transverse momentum dependence of elliptic flow
at different collisions energies.
The integrated elliptic flow is higher at RHIC than at SPS. 
The question is if $v_2$ as
a function of \pt~become steeper or if the observed increase 
in integrated flow is mostly due to the increase in mean transverse momentum?  
To answer this very interesting question is more difficult, as the
slope of $v_2(p_t)$ strongly depends on centrality and one does have
to have the same centralities in different experiments for 
comparison.
With the existing data, however, I would conclude that 
in the {\em central} collisions the slope of $\vtpt$ 
at SPS energies looks significantly smaller 
than at RHIC; 
in {\em mid-central} collisions the difference is not that large. 
(I discuss the centrality dependence of $v_2$ in more detail below.)

\subsection{Flow and  non-flow from multi-particle correlations}

Anisotropic flow is a multi-particle phenomena. 
It means that if one considers many-particle correlations 
instead of two-particle correlations, the relative contribution of non-flow    
effects (due to few particle clusters) would decrease.  Considering    
many-particle correlations, one has to subtract the contribution from    
correlations of the lower-order multiplets and
use cumulants instead of simple correlators~\cite{olli-cumulants}.  
For example, correlating four particles, one gets ($u_i \equiv e^{i\phi_i}$)   
\be    
\la u_{n,1} u_{n,2} u_{n,3}^* u_{n,4}^* \ra    
= v_n^4 + 2 \cdot 2 \cdot v_n^2 \delta_n + 2 \delta_n^2 \,,    
\label{fournf}    
\ee    
where $\delta_n$ stands for 
the non-flow contribution to two particle correlator: 
$\la u_{n,1}  u_{n,2}^* \ra = v_n^2+\delta_n$.
Subtracting from the expression (\ref{fournf}) twice the    
square of the two particle correlator, one is left with only the     
flow contributions     
\be    
\la \la  u_{n,1} u_{n,2} u_{n,3}^* u_{n,4}^* \ra \ra        
\equiv    
\la u_{n,1} u_{n,2} u_{n,3}^* u_{n,4}^* \ra    
-2 \la u_{n,1} u_{n,2}^* \ra ^2    
= -v_n^4 \,,    
\ee
where the notation $\la \la ... \ra \ra$ is used to denote
the {\em  cumulant}.  
A very elegant way of calculating cumulants in flow analysis with the
help of the generating function was proposed in~\cite{olli-cumulants}.
The NA49 Collaboration presented elliptic flow results from 4-, 6- and
even 8-particle cumulants~\cite{borghini}. 
Remarkably, the results from all higher
order cumulants agree well, indicating that even 4-particle cumulants
already suppress the non-flow contribution to a negligible level.
The use of the multi-particle correlations permits one to
measure the non-flow contribution and check its multiplicity
dependence.
The latter is expected to be roughly inversely proportional to the
multiplicity, thus reflecting the small cluster origin of such
correlations. Indeed, the result reported by the NA49 Collaboration are
in agreement with a $1/N$ dependence. 

The STAR Collaboration presented  results from 4-particle correlations
at \snn=130~GeV and \snn=200~GeV~\cite{ray}, Fig.~2.
Transverse momentum dependence of the non-flow (relative) contribution was
also discussed~\cite{filimonov}. The results are consistent with
weak increase in the relative non-flow signal from 15\% 
at low $p_t$ to about 20\% at $p_t\sim4$~GeV/c (at \snn=130~GeV). 
Finally, note a possible increase of the relative non-flow
contribution with the collision energy, 
from 7-10\% at SPS~\cite{borghini} to about 15\% at \snn=130~GeV 
and $\sim$20\% at \snn = 200~GeV at RHIC~\cite{filimonov}. 
Both effects are consistent with the increase in the contribution 
from hard processes.

\subsubsection{Flow fluctuations}

Very good statistics results reported at this conference become sensitive   
to another effect usually neglected in flow analysis, namely,    
event-by-event {\em flow fluctuations}.  The latter can can be due to two   
different reasons: ``real'' flow fluctuations -- fluctuations    
at fixed impact parameter and fixed multiplicity (see, for example,   
~\cite{kodama}), and impact parameter variations among events from    
the same centrality bin in a case where flow does not fluctuate at fixed    
impact parameter.     
Note that these fluctuations affect
any kind of analysis, including the ``standard'' one based on   
pair correlations.  The reason is that any flow measurement is    
based on correlations between particles, which     
are sensitive only to certain moments of the distribution in $v_2$.   
In the pair correlation approach with the reaction plane determined    
from the second harmonic, the correlations are proportional   
to $v^2$.  Averaging over events gives $\la v^2 \ra$,   
which in general is not equal to $\la v \ra^2$.     
The 4-particle cumulant method involves the   
difference between 4-particle correlations and (twice) the square of the   
2-particle correlations.  
It is usually {\em assumed} that this   
difference comes from non-flow correlations.    
Note, however, that this    
difference ($\la v^4 \ra - \la v^2 \ra^2 \ne 0$) could be   
due to flow fluctuations.  Let us consider an example where the   
distribution in $v$ is flat from $v=0$ to $v=v_{\rm max}$ and there is
no non-flow contribution.   
Then, a simple calculation would lead to the ratio of the flow values   
from the standard 2-particle correlation method and 4-particle cumulants   
as large as $\la v^2 \ra^{1/2}/(2\la v^2 \ra^2 - \la v^4 \ra)^{1/4}   
= 5^{1/4}\approx 1.5$.   
Thus comparing any theoretical model with the data one should take
into account flow fluctuations if they exist in the model.
\vspace*{-7mm}
\begin{figure}[htb]
\begin{minipage}[t]{78mm}
\includegraphics[height=6cm,width=78mm]{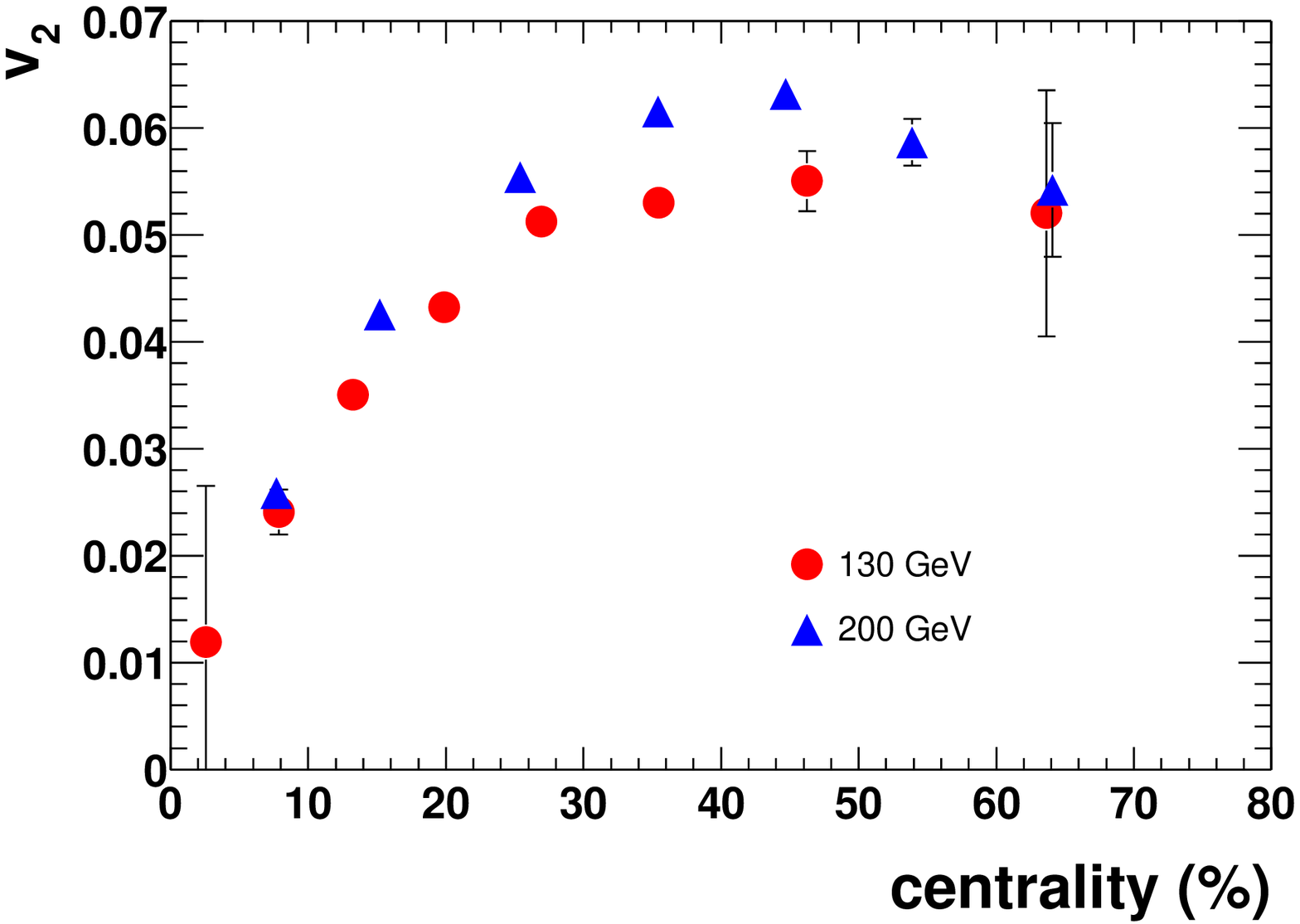}
\vspace*{-9mm}
\caption{STAR Preliminary results~\cite{ray} 
on centrality dependence of elliptic
flow from 4-particle cumulant analysis.}
\label{fv2cent}
\end{minipage}
\hspace{\fill}
\begin{minipage}[t]{78mm}
\includegraphics[height=6.5cm,width=79mm]{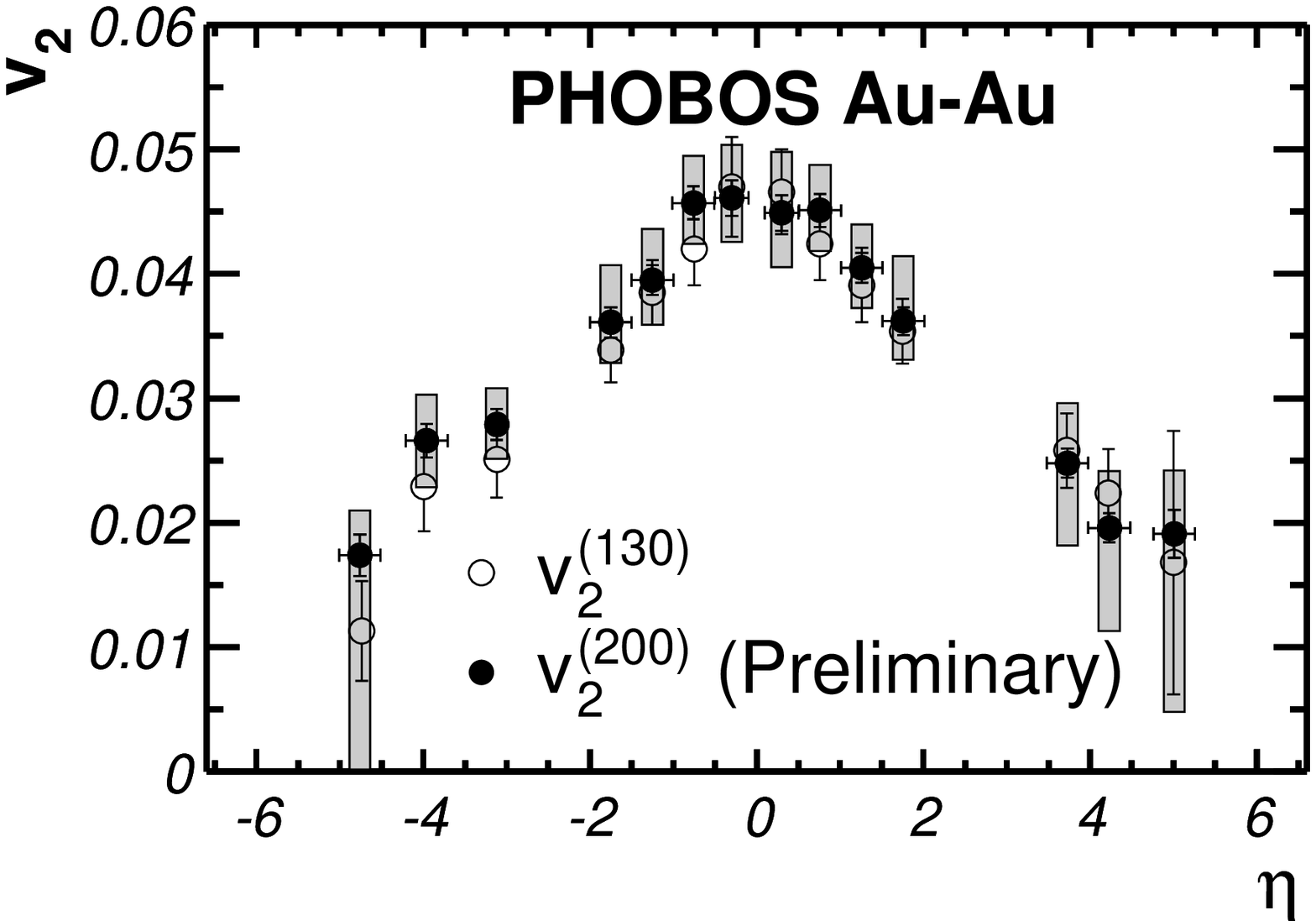}
\vspace*{-9mm}
\caption{PHOBOS preliminary results~\cite{manly} 
on pseudorapidity dependence of
elliptic flow at \snn=200~GeV.}
\label{fv2eta}
\end{minipage}
\end{figure}

\vspace*{-4mm}
\subsection{Pseudorapidity dependence}

The PHOBOS Collaboration is still the only one which reports elliptic
flow in a wide pseudorapidity range, Fig.~\ref{fv2eta}. 
This pseudorapidity dependence is still a challenge for the
hydrodynamic models~\cite{hirano}, which predicts rather weak
pseudorapidity dependence of elliptic flow. As was pointed out at the
conference such behavior of $v_2$ in the hydrodynamic model could be
related to the boost invariant initial conditions used in the calculations.  
Note also that PHOBOS results for
elliptic flow at \snn=130~GeV and 200~GeV are very similar, which is
in some disagreement with the STAR results.

\subsection{Elliptic flow at large $p_t$}

An obvious interest in elliptic flow at large transverse momenta
is due to a possible strong signal as a results of high $p_t$
parton energy loss~\cite{snellings,gyulassy}. 
Shuryak in his recent paper ~\cite{shuryakv2max} 
argued that the signal observed by the STAR Collaboration
(only elliptic flow results from
2-particle correlations were known at that moment)
could be probably {\em too} strong.
He has calculated elliptic flow for the limiting case of very strong
absorption resulting in predominant surface emission. 
Using his idea I arrived for  the maximum flow
values as a function of impact parameter 
\be
v_{2,max}=\sin(2\alpha)/(6\alpha),  
\; \; \; \; \; \; \; \; \; \; \; \; 
{\rm where}\;\; \alpha=\arccos(b/(2R)).
\ee
Comparing this formula to the STAR data~\cite{star_cumulants} from
4-particle cumulants one finds reasonable agreement.

At this conference, the STAR Collaboration presented 
the results~\cite{filimonov} on $v_2(p_t)$ up to
$p_t\approx 12$~GeV/c (from 2-particle correlations). 
It shows that elliptic flow is large at least
to $p_t\approx 6 \;-\; 8$~GeV/c.
Elliptic flow is maximum at $p_t \sim 3$~GeV/c 
and probably decreases very slowly up to
the highest momentum measured. 
To discuss reliably any trends in the region of $p_t> 6$~GeV/c,
however, requires multi-particle correlation measurements.
Model calculations are also needed. 

Another interesting fact concerning flow at high $p_t$ has been
reported by the CERES Collaboration~\cite{slivova}.
According to their data, the saturation of $v_2(p_t)$ 
at SPS energies is quite similar to the one at RHIC, although
to make a definite conclusion one needs more carefully comparison.

\subsection{Baryon and meson elliptic flow}

At this conference both the STAR and PHENIX Collaborations 
presented very interesting results on \pt dependence of elliptic 
flow of baryons and mesons. 
Although the results from both experiments have rather large
error-bars, both agree that baryon elliptic flow saturates 
at higher \pt and larger $v_2$ values than meson elliptic flow. 
From my point of view it could be due to the (partial) particle production via
(constituent) quark coalescence at moderate transverse momentum. 
Then, in this region:
\be
\frac{d^3n_M}{d^3p_M} \propto [\frac{d^3n_q}{d^3p_q}(p_q\approx p_M/2)]^2,
\; \; \; \; \; \; \; \; 
{\rm and}\;\; \;\;
\frac{d^3n_B}{d^3p_B} \propto [\frac{d^3n_q}{d^3p_q}(p_q\approx p_B/3)]^3,
\label{e:coales}
\ee
resulting in larger saturation values of $v_2$ for baryons as
well as the saturation point being at higher transverse momentum.

Note that Eq.~\ref{e:coales} is
valid only in a rather limited region of transverse momentum.
At very low momentum probably most hadrons are formed via coalescence.
Then their elliptic flow values should be similar to the elliptic flow 
of constituent quarks (I do not consider here the possible dependence
on the mass of the particles). 
At really high $p_t$, however,
hadrons are formed mostly due to parton fragmentation. There, 
the elliptic flow of hadrons should be determined by the elliptic flow
of the fragmenting partons and there should be no significant difference in
baryon and meson flow (due to this mechanism and neglecting the
difference in the parton fragmentation functions into baryons/mesons). 
Note that the coalescence effect, if present, would also lead to an increase of
the relative baryon to meson ratio in the same $p_t$ range. 

The centrality dependence of the ratio 
$v_{2,baryon}(p_t)/v_{2,meson}(p_t)$ would be extremely interesting.
If the effect of baryon flow being larger than meson flow  
is indeed due to coalescence, one would expect that
at higher centralities the crossings would appear at somewhat higher
transverse momentum.
A detailed comparison of the transverse momentum and centrality
dependencies of elliptic flow of baryons/mesons and their yields could
also help to understand the nature of both effects.

\subsection{Centrality dependence}

In the hydrodynamic limit elliptic flow is approximately
proportional to the original spatial ellipticity of the
nuclear overlapping region~\cite{Olli92,kolb,Shuryak99}, $v_2\propto \eps$.
In the opposite limit, 
usually called the {\em low density limit}~\cite{Heiselberg,vpPLB},
elliptic flow depends also on the particle density 
in the transverse plane: $v_2\propto \eps \;dN/dy \;/S$, where $S$ is the
area of the overlapping zone. 
The comparison of the results on elliptic flow from the point of view
of particle density in the transverse plane 
was first done in~\cite{vpPLB}. 
In this approach, the transition 
to deconfinement would lead to some wiggles in $v_2/\eps$ dependence,
(``kinks'')~\cite{Sorge99,Heiselberg,vpPLB}). 
\begin{figure}
\centerline{\includegraphics[height=12.cm,width=14cm]{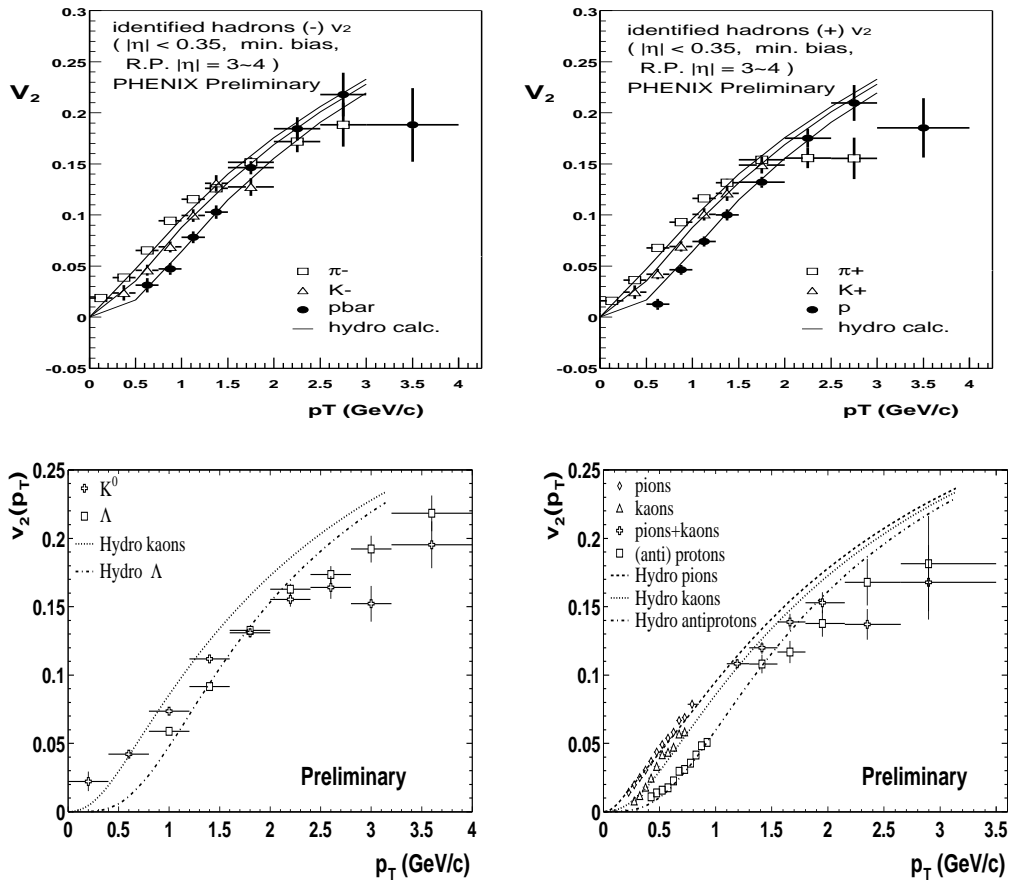}}
\vspace*{-9mm}
\caption{
Upper panels: PHENIX Preliminary data on elliptic flow of
protons/anti-protons and pions. 
Lower panels: STAR preliminary data
on elliptic flow of $\Lambda$'s and $K^0$ and protons and pions.
}
\label{fBM}
\end{figure}

\begin{figure}[htb]
\vspace*{-9mm}
\begin{minipage}[t]{84mm}
\includegraphics[height=7cm]{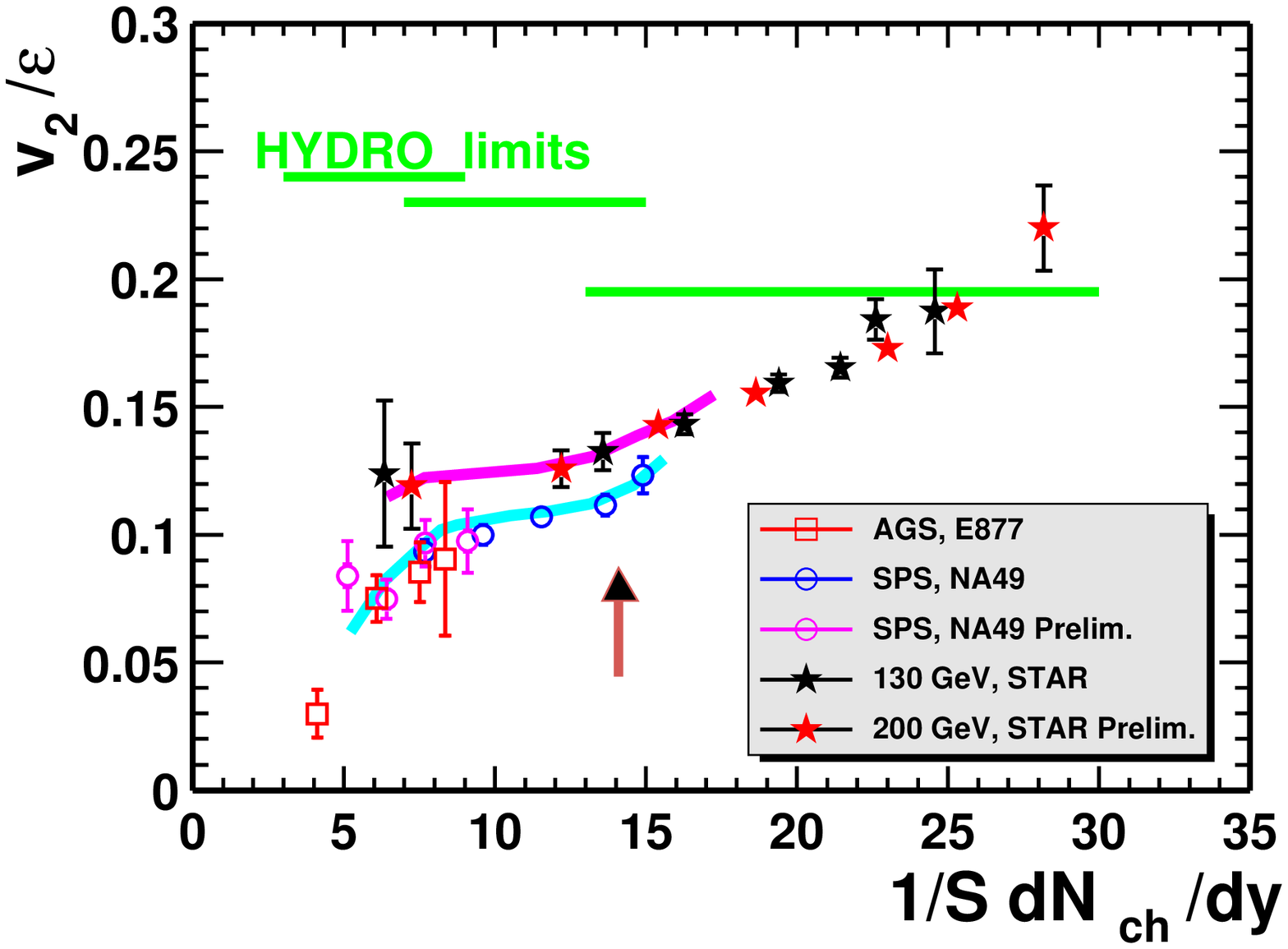}
\vspace*{-15mm}
\caption{$v_2/\eps$ as a function of particle density. The curves are
only to indicate the possible structure.}
\label{fscaling}
\end{minipage}
\hspace{\fill}
\begin{minipage}[t]{71mm}
\includegraphics[height=7.cm]{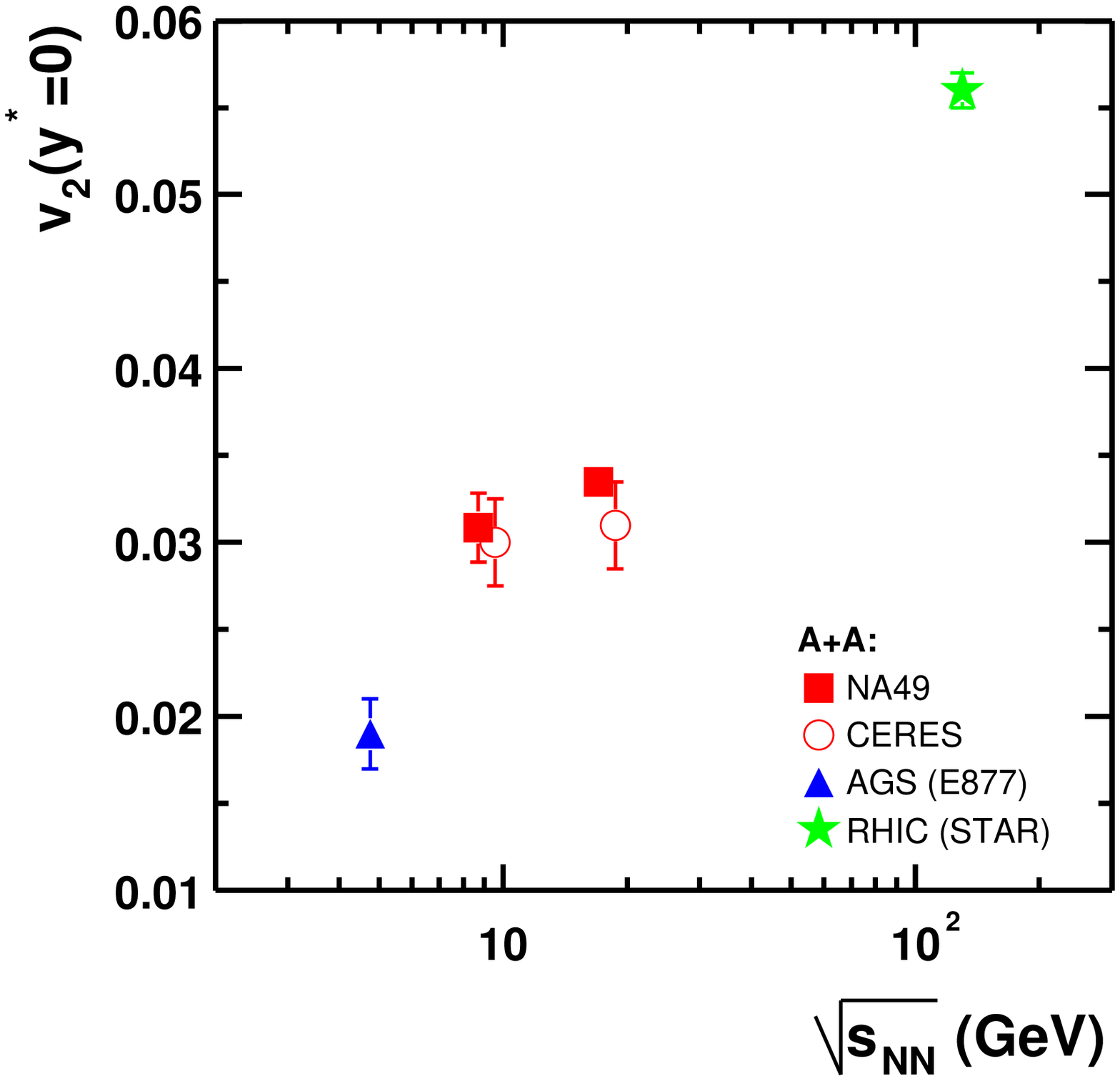}
\vspace*{-15mm}
\caption{Pion elliptic flow at midrapidity for 25\% centrality versus
\snn}
\label{fv2edep}
\end{minipage}
\end{figure}

Fig.~\ref{fscaling} shows the recent ($\sqrt{s_{NN}}=130$~GeV) STAR results 
on elliptic flow from 4-particle cumulants~\cite{star_cumulants},
preliminary 
NA49~\cite{wetzler} (also from 4-particle cumulants), 
and E877 results.
New measurements at full RHIC energies of $\sqrt{s_{NN}}=200$~GeV
reported by STAR~\cite{ray} are also shown. 
For this Figure I have rescaled down these new data points by 
a factor of 1.06 to account for the change in mean transverse 
momentum (in 200 GeV data
set the low transverse momentum cutoff is 150 MeV/c compared to 75 MeV
in 130 GeV data). 
Note also that SPS data do not have a low $p_t$ cutoff.
(There are  possible systematic
errors in Fig.~8 of the order of 10-20\% due to uncertainties 
in centrality measurements. However, these uncertainties do not alter the
general trend).   
Fig.~\ref{fscaling} shows that at high particle densities the data
reach hydrodynamical limits. With relatively large error-bars for the
most central collisions it is difficult to draw any conclusion about
the saturation at the hydrodynamical limit. 
As for the hydrodynamic limits, they are also somewhat model dependent.
In some cases (depending on the latent heat and equation of
state) in the model~\cite{Shuryak99} one can observe an almost
linear dependence of $v_2$ on particle density.

A closer examination of the plot in the region of 7-15 particles/fm$^2$
shows that there is no scaling in this region. 
(This is probably related to the use of $\eps$ calculated using the  specific 
weight according to the number of wounded nucleons. One should
also take into account slightly different transverse momentum regions
used in different analyses.) 
Note that the data from both RHIC and SPS results show a rather 
flat region with some rise at density \~15 fm$^{-2}$. 
One could speculate that such a behavior
could indicate a change in the physics of rescatterings. 
Noteworthy that the color percolation point discussed by
Satz~\cite{satz} is just in the same region (shown by an arrow in
figure).
Very interesting that the WA98 Collaboration~\cite{bathe} analyzing the high
transverse momentum photon flow also observes a non-smooth behavior
at the same place.

Another indication of non smooth flow dependence on the particle
density in this region can be seen in the figure showing
an elliptic flow excitation function~\cite{wetzler}.
In this figure both NA49 and NA45 data points show very little
increase in the average flow from 40$A$~GeV to 160$A$~GeV  data, while
there is a significant increase between highest AGS point and lowest SPS,
and between SPS and RHIC.\footnote{In the poster of H.~Sako, CERES
Collaboration,
one can also notice an interesting increase in multiplicity as well as
mean transverse momentum fluctuations at beam energy of 40$A$ GeV.}

\subsection{Models}

It has been mentioned many times that at RHIC energies the
hydrodynamic description of the results is quite successful.
However, a few problems, like the description of the pseudorapidity
dependence~\cite{hirano} still remain.
The hydrodynamic model has been discussed in a separate
talk~\cite{pasi} at this conference
and many details can be found there.

At this conference there also was presented a model~\cite{humanic}
with totally hadronic rescatterings starting at zero time, 
which also describes many  features of the data.

AMPT (A Multi Phase Transport) model~\cite{lin}
also describes the data very well in the so-called ``string melting''
scenario with quark-quark cross sections of about 5~mb. More detailed
examination of the ``string melting'' scenario reveals that it is
nothing else than what one would do in a case of a constituent quark
picture. Namely, the number of quarks in the system is calculated as
the number of quarks in the final hadrons. The final formation time
(after which a quark can interact) is also taken into account. The
cross section of 5~mb also does not look unusual - it would be close
to the total NN cross section divided by 9. 
Note that the AMPT model also quite successfully describes the HBT radii.
This model definitely deserves further study.

\section{Conclusions}

I will summarize the most important results just by enumeration:
\\
- the ``wiggle'' in proton directed flow has been observed at 
SPS~\cite{wetzler}.
\\
- non-flow contribution is better understood and {\em measured} both
at SPS and RHIC~\cite{borghini,ray}.
\\
- at RHIC $v_2(p_t)$ saturates at values close the ones in 
the surface emission picture.
\\
- $v_2$ of baryons exceeds $v_2$ of mesons from  $p_t$ about 2 GeV/c
till at least 3--4~GeV/c~\cite{esumi,filimonov}.
\\
- At RHIC elliptic flow persists up to $p_t \approx 8-12$ GeV/c
\\
- the saturation of elliptic flow at SPS probably not very different
than at RHIC~\cite{slivova}
\\
- The dependence of elliptic flow on pseudorapidity is rather strong.
It drops about 2 times at about $\eta=3$~\cite{manly}. 
(Still a challenge for the hydrodynamic description.)
\\
- $v_2/\eps$ is roughly proportional to particle density in the
transverse plane $(dN/dy)/S$ reaching the hydrodynamic model
predictions in central collisions.\\

In this summary I also want to mention  a few important
measurements for the future:\\
- we have to measure non-flow correlations in $pp$ collisions and
compare them to the ones measured in nuclear collisions.
 \\
- we need to check for non-flow contribution at high transverse momenta.
 \\
- centrality dependence of baryon/meson  elliptic flow would be very 
interesting.
\\
- the collision of lighter system (e.g. Cu+Cu)  
and (multi)strange particle flow will be very important to understand
the origin of the observed collectivity.
\\ 
- different RHIC collaborations have data taken at \snn=20~GeV. It
would be really interesting to analyze these data and compare with
SPS results.
\\ 
- directed flow at RHIC will be probably the next result from RHIC... 
\\
- the percolation point region should be looked at in great detail.
\\
- finally, we need a  detailed comparison (at a level
of a few percent) of the results from different experiments; for that
we need the results corresponding to the same centralities.

{\bf Acknowledgments.}
I thank the organizers for the invitation, 
all the speakers for their help providing me their data, and many of
my colleagues, and in particular Art Poskanzer, for numerous very helpful
discussions.


\end{document}